# Overview of Recent Studies and Design Changes for the FNAL Magnetron Ion Source

D.S. Bollinger[1, a)] and A. Sosa[1, b)]

[1]Fermi National Accelerator Laboratory, P.O. Box 500, Batavia, Illinois, 605010
[a)]Corresponding author: bollinger@fnal.gov
[b)]asosa@fnal.gov

**Abstract.** This paper will cover several studies and design changes that will eventually be implemented to the Fermi National Accelerator Laboratory (FNAL) magnetron ion source. The topics include tungsten cathode insert, solenoid gas valves, current controlled arc pulser, cesium boiler redesign, gas mixtures of hydrogen and nitrogen, and duty factor reduction. The studies were performed on the FNAL test stand described in [1], with the aim to improve source lifetime, stability, and reducing the amount of tuning needed.

## INTRODUCTION

The FNAL magnetron ion sources provide ions for the neutrino program and other experiments. The typical run time in recent years has been at least 9 months long. As a result the ion sources need to provide stable beams for long periods of time. The studies and design changes discussed in this paper were performed to meet this goal. Issues such as hydrogen gas pressure and cesium boiler temperature greatly affect the stability of the source. And cathode erosion, which contributes to lifetime, is in part due to the duty factor seen by the ion source and the choice of cathode material.

## TUNGSTEN CATHODE INSERT

One of the aging mechanisms of magnetron ion sources is cathode erosion. The erosion is caused by back scattered positive ions, which can cause the cathode to lose material that ends up restricting apertures such as the hydrogen and cesium inlets and the anode extraction aperture. Also, the cathode erosion typically occurs in the center of the spherical dimple, changing the focus of the dimple. One possibility to reduce this erosion is to use a harder cathode material. Current magnetron ion sources use cesiated molybdenum (Mo) cathodes which have a minimum work function of about 1.6 eV and are easily machined into the proper shape. Tungsten (W) a harder material, has a somewhat higher minimum work function and is more difficult to machine but, may be less susceptible to erosion. A molybdenum cathode, machined to accept a tungsten insert provided by V. Dudnikov and suggested in [2], was installed in a source and studied in our test stand. Figure 1 shows the Mo cathode with the W insert, which was a press fit.

One difference between molybdenum and tungsten is that the Cs/W work function does not achieve the lower value of Cs/Mo, but the H- production peak is broader, as shown in Figure 8.10, page 341 of reference [3]. If the H– production peak is broader the Cs coverage may be more forgiving allowing for larger changes in the surface work function while still producing enough H− ions.  One way to change the surface work function of the cathode is to change the thickness of the Cs layer on the cathode. Two of the parameters that affect the Cs coverage are the duty factor and cathode temperature. A study was performed where the arc pulse width was changed while maintaining the same arc current and Cs boiler temperature. Perveance scans were taken at different duty factors and cathode temperatures for both the Mo and W insert cathodes.

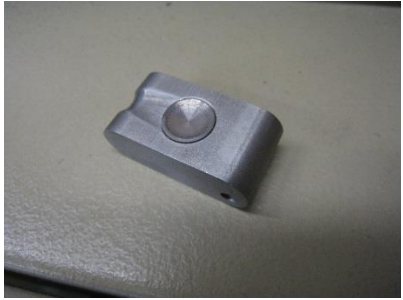

**FIGURE 1**. Molybdenum cathode with tungsten insert, which was installed in a magnetron to be studied in our test stand. The insert was provided by V. Dudnikov and was press fit into a specially machined Mo cathode.

Figure 2a shows the effect of different duty factors on the extracted beam current at 35 kV. The beam current was lower for the W cathode which was expected, however difference in the shape of the curve between the two cathodes is inconclusive. Figure 2b shows the beam current as a function of the cathode temperature, which should correlate to the cesium layer thickness. Once again the beam current is lower for W, but the shape of the W curve does seem broader. We do plan on taking more data with the W cathode and possibly install in an operational system to see how the much erosion is visible in the center of the dimple, from back scattered positive ions over a period of several months.

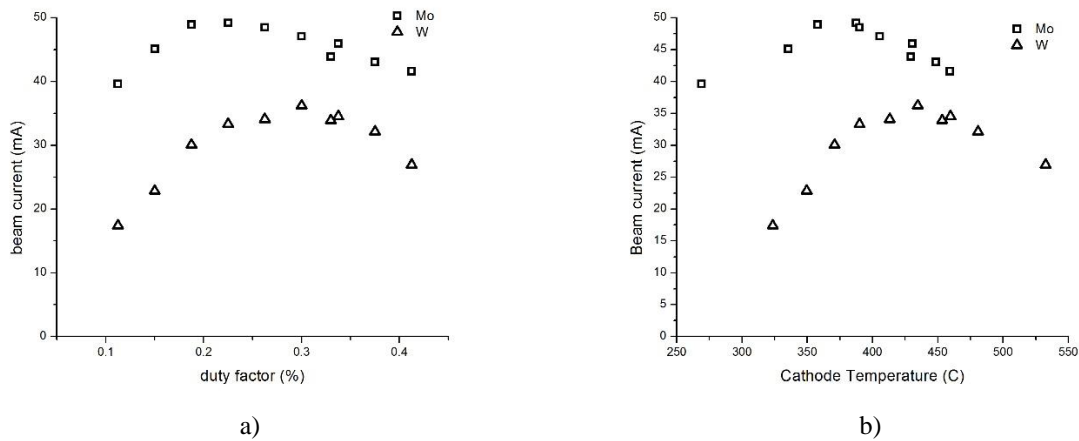

a)          b)

**FIGURE 2**. a) Difference in beam current at 35 kV for different arc pulse widths for Mo and W cathodes. b) Difference between Mo and W cathodes at 35 kV extraction with different cathode temperatures.

## SOLENOID GAS VALVE

The FNAL H− ion sources have used piezoelectric valves for injection of hydrogen into the source since they were originally designed [4]. These valves are very susceptible to changes in temperature. When sources operated in the Cockcroft-Walton accelerators a feedback loop based on average accelerating column vacuum pressure was used along with a temperature and humidity controlled room. The feedback loop relied on a modified ion gauge controller which had a specially designed card to reduce the noise on the controller analog output. With the installation of the RFQ, new ion sources and electronics, the ability to use a feedback were lost, and the room where the sources are installed is no longer temperature controlled. As a result the gas pressure inside the source is now anti-correlated to changes in the room temperature as can be seen in Fig. 3a. Changes in source pressure affect the arc and beam currents, as well as the shape of the plasma meniscus [5] which affect the losses and transmission efficiency of the linac. Other labs such as ISIS use a collar around the valve for cooling [6], however this would be difficult to implement due to our current extraction scheme.

A solenoid gas valve used on the Brookhaven National Laboratory BNL, Electron Beam Ion Source EBIS source [7], has been studied on the FNAL test stand [1]. Also a gas valve test stand was built for testing valves with an emphasis on comparing the piezoelectric and solenoid valves. However, finding an appropriate pressure sensor that was both fast enough to measure rise and fall times and sensitive enough to operate the pressures seen in the ion source vacuum has proven to be difficult.

One attempt to compare the solenoid to the piezoelectric valves was to operate the source with 320 V DC between the anode and cathode, while pulsing the gas valve. Preliminary results of this study are shown in Fig. 3b. Several scope traces were taken of the arc current using the same source with similar initial conditions, such as cathode and cesium boiler temperature and average vacuum pressure. In Fig. 3b, region one is the time between the gas valve opening and the beginning of the arc would be the amount of time that it takes for the internal source pressure to reach a value high enough for plasma generation. As the pressure continues to increase in region 2 the arc current would increase. Finally in region 3 the arc current would decrease as the hydrogen is pumped out of the source through the anode aperture until there is not enough pressure to sustain an arc. As can be seen in the plot the fill time of the source with the solenoid valve is faster, but the peak pressure is not quite as high. Further study will be needed to tell if this method is a meaningful way of characterizing gas valve performance.

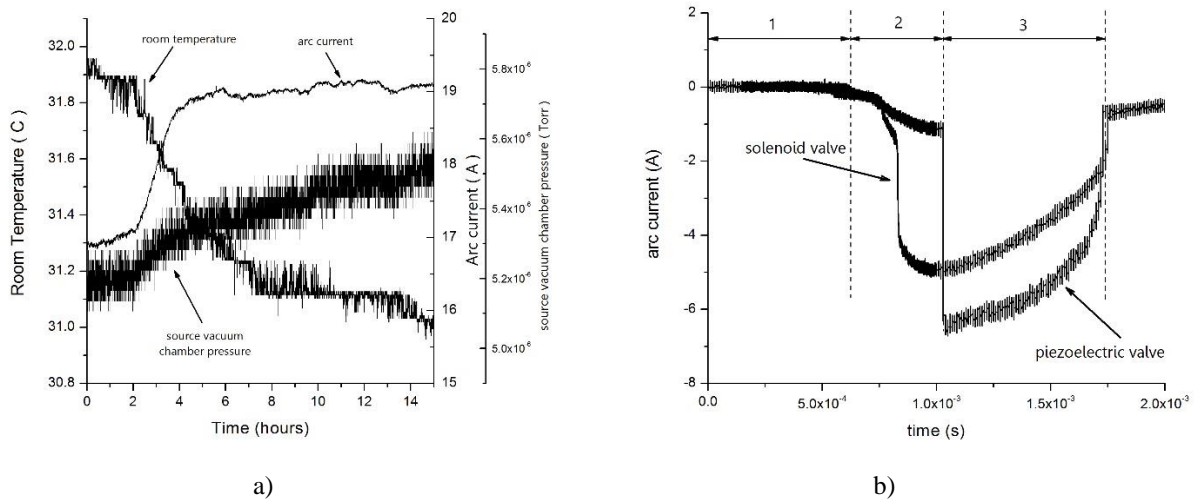

a)　　　　　　　　　　　　　　　　　　　　　　　　b)

**FIGURE 3**. a) Data logger plot showing the dependence of gas pressure in the source vacuum chamber with changes in room temperature. b) Comparison of the arc current while operating the source DC between the piezoelectric and solenoid valves

## CURRENT REGULATED ARC PULSER

The arc pulser which has been used since the installation of the Radio Frequency Quadrupole (RFQ) and direct extraction magnetron ion source [8] is a 300 V voltage regulated pulser. There is a clear relationship between arc current and extracted ion beam current as can be seen in Fig. 4a which is a typical perveance at several arc currents. In order to have stable beam current the arc supply voltage needs to be adjusted frequently as can be seen in Fig. 4b, which shows a nine month operational run. Feedback loops were written to adjust the arc supply voltage to keep the current constant. However, the loops could not handle rapid changes in arc current and they would tend to run away, so they were abandoned. Also attempts to modify the voltage regulated circuit did not work as expected.

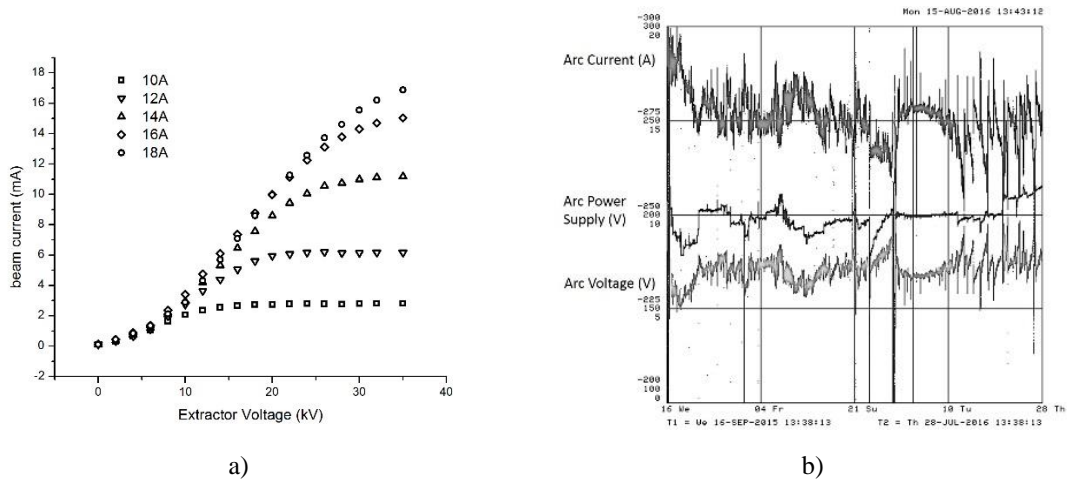

**FIUGRE 4**. a) Typical beam current preveance curve taken at different arc currents. b) Nine month long operational run showing variations in the arc current and voltage. The arc supply voltage is adjusted frequently to maintain enough arc current for suppling constant beam current for operations.

In order to help maintain a stable beam current, a current regulated modulator was designed and is being studied on the test stand [9]. Figure 5a shows several hours of running the source with the new modulator. As can be seen the arc current is constant and the arc voltage is changing as the source warms up, meaning that the source impedance is changing during this time. Figure 5b shows a weekend long run where the current varies by a 0.5 A. It is not entirely clear at this time why the regulation is ± 0.5 A. However, this variation is small compared to the voltage regulated modulator.

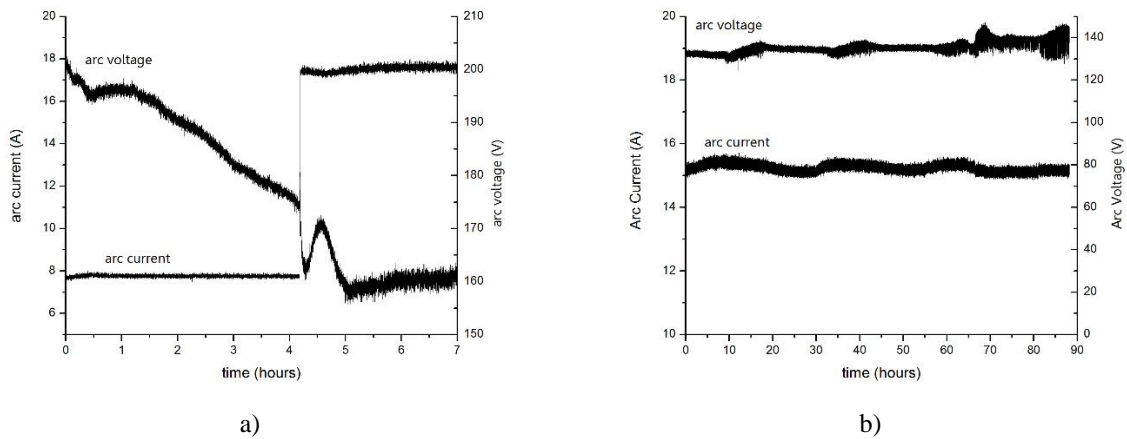

**FIGURE 5.** a) Several hour run with new arc modulator showing that the arc voltage decreasing as the arc current remains constant. b) Weekend long run with the same modulator showing that the arc current is constant to ± 0.5 A.

At this time extractor sparking has been causing the feedback circuit to occasionally fail. We plan to continue to improve the reliability and regulation of this pulser. Several questions remain how regulating current will affect source operations, such as how changes in source impedance will affect the beam quality and current, and will a more stable arc current lead to more stable source operation.

# CESIUM BOILER REDESIGN

The current cesium system consists of a copper tube boiler, stainless steel valve and cesium feed tube. The copper boiler is heated by a heat tape which also is wrapped around the valve and feed tube. The valve has a resistive heater wire wrapped around it and the feed tube has a resistive heater wire inside the tube. There are separate power supplies for each heater.

A five gram glass ampule of cesium is installed in the copper boiler. Once the system is installed on a source and pumped down, the tube is crimped to break the glass ampule and release the cesium. This method of releasing cesium has been used since the 1980's.

When the ion sources were used in the Cockcroft-Walton accelerators, the boiler temperature was typically 145 C. The driving force for this temperature was the need for arc currents of 50 A, which were required to deliver 50 mA of H− beam current at 18 kV extraction voltage.

The Cockcroft-Walton accelerators were replaced by a Radio Frequency Quadrupole (RFQ) [8] which requires an injection energy of 35 kV. A new ion source, based on the Brookhaven National Laboratory (BNL) source [9] was built and installed. The new source has improved focusing and a higher extraction voltage which allows the source to run with arc currents around 15 A and routinely delivering over 60 mA of H− beam current in the middle of the LEBT.

The lower arc current requires less cesium for H− production. Now the cesium boiler temperature is less than 120 C. Since the boiler is made out of copper which is a good conductor of heat, most of its heating comes from the valve heater. In order to get some control of the boiler temperature we had to thin out the insulation that is wrapped around the boiler. With less insulation the boiler temperature is now affected by the room temperature as shown in Fig. 6a. Also, the current in the heat tape is typically around 500 mA with a tuning range of only 10's of mA.

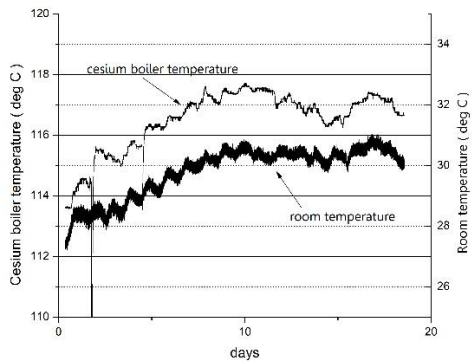 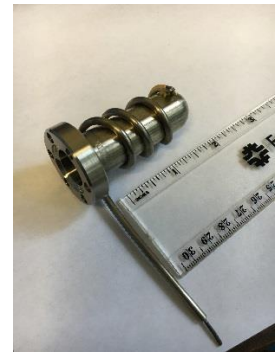

a) b)

**FIGURE 6**. a) Data logger plot showing the cesium boiler temperature following the room temperature due to the small amount of insulation on the boiler b) Picture of new stainless steel boiler with thick heater wire welded on.

A new stainless steel (SS) boiler using a heater wire, shown in Fig 6b, similar to one that BNL uses, was designed. Stainless steel was chosen because it is less thermally conductive than copper so, it would be less susceptible to heating by the valve. This should allow the installation of more insulation, to reduce the effect of changing room temperature, and with the use of a heater wire larger currents would be required for the same temperature giving more control of the boiler temperature. Figure 7a shows that the current required by the new SS boiler is an order of magnitude higher than the copper boiler with a wider tuning range. The SS boiler is also more power efficient as can be seen in Fig. 7b. For a given boiler temperature the SS boiler require between 13% and 15% less power, mainly due to the low resistance of the wire. We are planning on starting up for the next operational run with new SS boilers on both sources

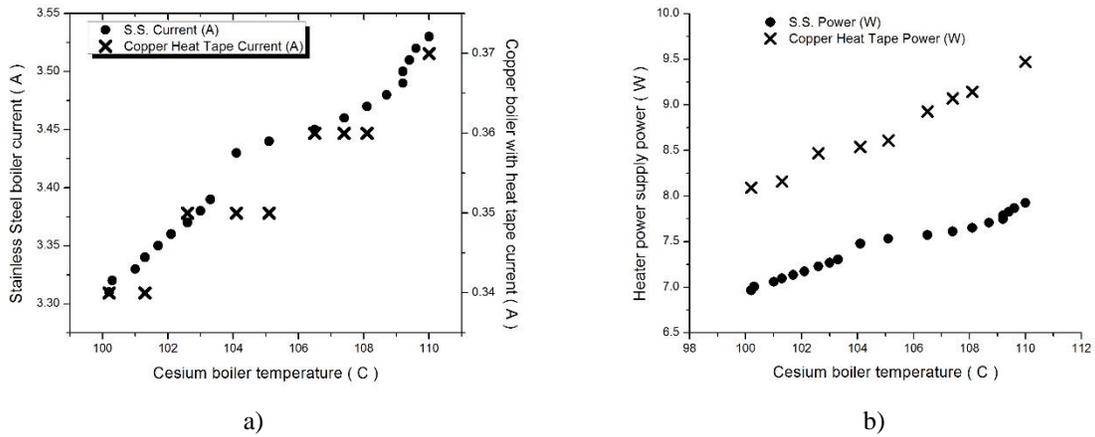

a)   b)

**FIGURE 7.** a) Boiler temperature as a function of heater power supply current. The heat tape current on the copper boiler required very little current for the normal operating boiler temperature of 110 C. b) Boiler temperature as a function of heater power. It is interesting that the stainless steel boiler takes much less power for a given temperature.

## GAS MIXTURE STUDIES

During single turn injection into the FNAL 8 GeV Booster ring, there is a bunch by bunch variation on the beam current [10]. One possibility is that the variation in beam current originates in the ion source and is preserved through the linac. Previous studies on magnetron beam current noise included cathode geometries and gas mixtures of hydrogen and nitrogen [8]. The gas mixture studies used hydrogen bottles which were purchased with different percentages of nitrogen mixed in. The study was inconclusive because the error in the analyzed accuracy of the mixture was larger than the percentage of nitrogen.

As a result of the previous study mass flow controllers were installed on the FNAL test stand as described in [1]. The controllers allow for fine control of gas mixtures and the mixing takes place just before the ion source gas valve. The studies involved three scans using different total mass flow (hydrogen flow + nitrogen flow) and mixtures of hydrogen with 0 % to 3.5% nitrogen. Figure 7 shows the noise ratio as a function of percent of nitrogen.

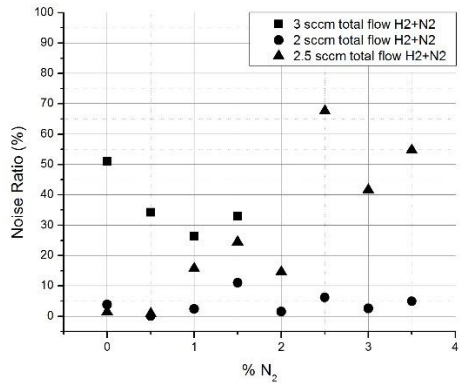

**FIUGRE 7.** Data from gas mixing studies taken a 30 kV extraction voltage shwoing the noise ratio for several total flow rates as a function of the percnet nitrogen added.

The noise ratio is the standard deviation of the amplitude divided by the amplitude of the toroid signal on an oscilloscope, with 30 kV extraction voltage. With 3 standard cubic centimeters per minute (sccm) of total flow the noise ratio was as much as 40% larger than with the lower flow rates and the source was not able to sustain a stable arc above 1.5% nitrogen. This is probably due to the source pressure being high enough that stripping of the H− ions was dominating. With lower flow rates the minimum noise ratio was seen at 0.5% nitrogen. However, the small signal on the toroid, often in millivolts, made the error large for these measurements. Even though this data looked promising, it is still inconclusive with the signal to noise that was seen. We are still pursuing the noisy source output and may revisit gas mixing in the near future

## DUTY FACTOR REDUCTION

The total arc current PW is determined by equation 1)

$$Total\ PW = Extractor\ rise\ time + LEBT\ neutralization\ time\ +\ Chopped\ beam\ PW \qquad 1)$$

The extractor rise time is factored in because during initial studies high spark rates occurred at the beginning of the arc pulse when the extractor was at full voltage. As a result the arc pulse starts prior to the extractor pulse. The original tube based extractor pulsers had a rise time of at least 130 µs plus an extra ~20 µs due to delays in the vacuum tube bias electronics. With a maximum chopped beam width of 62 µs, and observed neutralization time of approximately 30 µs [8], the required pulse width was 230 µs.

Recently solid state extractor pulsers described in [11] were installed on the Fermilab magnetron ion sources which have a rise time of 20 µs. This allows for the possibility of reducing the arc pulse width by at least 130 µs. And at a 15 Hz rep-rate, a reduction in duty factor of 43% would be realized. Since source lifetime is intimately tied to duty factor, studies were performed aimed to prove that stable operation at lower duty factor is possible.

A study in duty factor reduction was carried out on one of the operational sources. The arc pulse width reduced by 10 µs increments twice each day until the pulse width reached 100 µs. During this study the source was stable with a constant impedance, gas pressure and cesium boiler temperature. Figure 8a is a data logger plot of the reduction of the average arc power as a function of duty factor. The relationship is linear and the reduction in average arc power was 55% when the duty factor went from 0.30% down to 0.15%. This is a significant reduction in the average power in the source which should lead to longer lifetimes. Figure 8b shows the change in cathode temperature as a function of duty factor. Once again the relationship is linear and the change in temperature was 29% over the same duty factor span. With a lower cathode temperature, the cesium boiler temperature could be reduced and still maintain optimal cesium coverage, this should decrease the cesium consumption from the current $2.4 \frac{grams}{plasma\ day}$. We did begin to explore this prior to turning off the ion source. We do plan on restarting the source with a 100 µs arc PW and tuning for optimal beam current

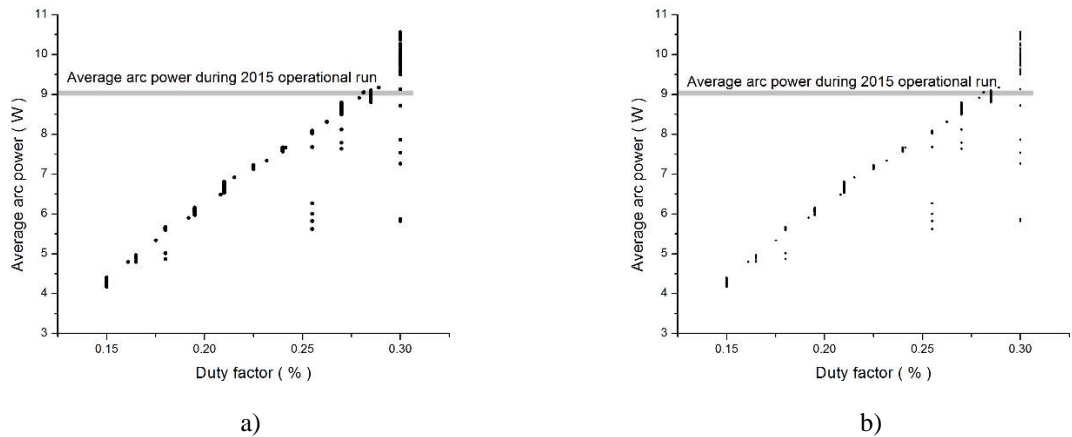

a)                                                                b)

**FIGURE 8.** a) Average arc power as a function of duty factor during time of duty factor reduction study. b) Cathode temperature as a function of duty factor during time of duty factor reduction study.

## CONCLUSIONS

The overview of studies presented in this paper are an ongoing effort to improve the reliability, lifetime, beam quality, stability and ease of operations of the FNAL magnetron ion sources. The longest lifetime to date for the ion sources has been 9 months of continuous operations. We were able to achieve this by frequent tuning of source parameters such as arc supply voltage, gas pressure and cesium boiler temperature. With the design of the current regulated arc pulser, implementation of solenoid gas valves and stainless steel cesium boiler the tuning required should be less. With less cathode erosion and lower cathode temperature from reduced duty factor the source lifetime should improve. We are still looking for ways to reduce the beam current variations. With improved diagnostics gas mixing could be studied in more detail to see if that is a viable way to reduce bunch intensity variations.

## ACKNOWLEDGMENTS

This research was supported by Fermi Research Alliance, LLC under Contract No. De-AC02-07CH11359 with the United States Department of Energy. The authors would like to thank Andrew Feld and Ken Koch for all of their work building the new ion source test stand and all of the modifications to the electronics and ion sources during the course of these studies. We would also like to thank C.Y. Tan for the design of the current feedback circuit and all of his help troubleshooting the modulator and for the many stimulating conversations about all of the topics presented in this paper.
.